\documentclass[fleqn]{article}
\usepackage[a4paper,margin=25.4mm]{geometry}
\usepackage[english]{babel}

\usepackage{amsmath}
\usepackage{amssymb}
\usepackage{sistyle}
\usepackage{cite}

\usepackage{tabularx}
\usepackage{graphicx}			
\usepackage[
	labelfont={sf,bf},			
	textfont={sf},			
	font=small,					
	justification=RaggedRight	
	]{caption}
\usepackage{subcaption}			
\usepackage{url}
\usepackage{xcolor}

\usepackage[unicode=true,pdfusetitle, bookmarks=false, breaklinks=false,pdfborder={0 0 0},backref=false,colorlinks=false]{hyperref}

\newcommand{\mean}[1]{\ensuremath{\mathrm{mean}\!\left(#1\right)}}
\newcommand{\var}[1]{\ensuremath{\mathrm{var}\!\left(#1\right)}}
\newcommand{\N}[1]{\ensuremath{\mathrm{N}\!\left(#1\right)}}

\newcommand{\LEDsym}{\blacktriangleright\;\!\!\!|}

\begin{document}

\begin{center}
	\hfill\\[1.0cm]
	{\Large\textbf{Direct measurement of darkness using a standard single-photon avalanche photodiode}}\\[0.2cm]
	{\large{T.H.A. van der Reep$\left.^{1,*}\right.$, 
	D. Molenaar$\left.^{2}\right.$ and
	W. L\"offler$\left.^{1}\right.$}} \\[0.2cm]
		{$\left.^{1}\right.$\emph{Leiden Institute of Physics, Niels Bohrweg $2$, $2333$ CA Leiden, The Netherlands}}\\[0.1cm]
	{$\left.^{2}\right.$\emph{Psychology Research Institute, Nieuwe Achtergracht $129$b, $1018$ WS Amsterdam, The Netherlands}}\\[0.1cm]

	$\left.^{*}\right.$reep@physics.leidenuniv.nl\\[0.2cm]
	\today
\end{center}
\begin{abstract}
	In experiments requiring extreme darkness, such as experiments probing the limits of human vision, assessment of the background photon flux is essential. However, direct measurement thereof with standard single photon detectors is challenged by dark counts and their fluctuations. Here we report an experiment and detailed statistical analysis of a direct measurement of darkness in a dedicated dark chamber suitable for human vision experiments, only using a standard single photon detector and a mechanical shutter. From a Bayesian analysis of $\SI{616}{h}$ of data, we find substantial to decisive evidence for absolute darkness (depending on choice of prior distribution) based on the Savage-Dickey ratio, and a light level $<\SI{0.039}{cnt/s}$ (posterior $0.95$-highest density interval).
\end{abstract}

\section{Introduction}\label{secIntroduction}
In psychophysics, the study of human perception, one of the parameters of interest is the smallest signal creating conscious perception. For the human visual system, it has been shown that photopic (colour) vision mediated by cone photoreceptor cells is sensitive to photon states with order $100$ photons \cite{KoenigHofer2011}. On the other hand, scotopic (low-light) vision using rod cells has been shown to be sensitive to few-photon states \cite{Hechtetal1941,Sakitt1972,Teichetal1982} using sources of Poissonian light. By using a source of spontaneous down-converted (SPDC) light it was found that humans might even perceive single photons \cite{Tinsleyetal2016}. Given the noisy environment of the brain, and the low efficiency of the eye, this is quite remarkable. To increase our understanding human vision in the few-photon limit further, we have recently proposed to study the perception of one- and few-photon states using quantum detector tomography \cite{Reepetal2023}.\\ 
In such studies, a dark chamber is of utmost importance. A common approach is to construct a dark chamber within a dark lab to null background light that would add to experimental noise, which, apart from arising from a possible background, mainly originates from spontaneous activation of the photoreceptors. In rod cells, e.g., the number of dark activations is $(3.4\pm\SI{0.8)e-3}{s^{-1}}$ \cite{Fieldetal2019}, implying approximately $50$ dark activations per second if the experiment is performed using a Maxwellian view (light pulses focussed on the eye lens \cite{Westheimer1966}), in which the light pulses are directed to the area of the retina where the rod cell density is highest ($\SI{1.6e5}{rods/mm^2}$ \cite{Cursioetal1990}) with a half-opening angle of $0.4^{\circ}$ and assuming a lens-retina distance of $\SI{24}{mm}$. These dark activations directly compete with the weak few-photon signals to be perceived, and this task becomes even harder if any background light is present. However, to the best of our knowledge, measurements of the actual level of darkness during these experiments and other studies on human visual perception at low light-level \cite{DentonPirenne1954,HalettMarriottRodger1962,
Holmesetal2017,Deyetal2021} are typically not reported.\\
 
Assessment of darkness to levels below $\SI{0.1}{photons/s/mm^2}$ is not easy due to dark counts in detection systems. For instance, standard single photon detectors show $10-1000$ dark counts per second. Therefore, one might attempt to measure the level of darkness of the lab environment, measure the attenuation of the dark chamber using a bright light source, and take the darkness level as the quotient of those values. Although this gives an estimate on how dark the dark chamber is, some important interpretation difficulties arise, especially when the goal is to measure absolute darkness. Firstly, since the light source used in this ``environment-attenuation'' (EA) approach is spatially and spectrally different from the actual light sources that give rise to background light in the environment, one only \emph{estimates} the attenuation of the dark chamber. Secondly, and more importantly, this method does not take into account light generated \emph{inside} the dark chamber, e.g., due to fluorescence, phosphorescence or even low levels of radio-activity in the chambers' walls.\\
These issues can be overcome by using a \emph{direct} measurement of the level of darkness, without using additional sources of light and from within the dark chamber itself. To this end we place a single-photon detector in our dark chamber and connect it to a shutter. As such, we can study differences in the detector's count distribution, dependent on whether the shutter is open or closed. In section \ref{secMethod} we elaborate on the general method and describe the experimental setup of the dark lab and dark chamber, and in section \ref{secBayes} the method is applied to assess the level of darkness in our dark chamber. We analyse the data using a Bayesian approach which, contrary to frequentist approaches based on $p$-values \cite{Wagenmakersetal2008}, can be used to quantify both the presence and absence of evidence supporting the null-hypothesis of absolute darkness. Our results from using the EA approach are presented in section \ref{secBright}, after which we provide a general discussion and conclusion in sections \ref{secDiscussion} and \ref{secConclusions} respectively.

\section{Method}\label{secMethod}
In order to test for absolute darkness we use the following method. First, we darken the lab by applying sheets of $\SI{5}{mm}$-thick black foam board\footnote{Budget foamboard $\SI{5}{mm}\ 70\times 100$ zwart} to all lab windows, which we fixate to each other and the window frame with $\SI{50}{\mu m}$-thick aluminium tape. A layer of black neoprene\footnote{Celrubberplaat EPDM zk $\SI{2}{mm}$} with a thickness of $\SI{2}{mm}$ is applied as weatherstrip to the inside of the lab door frame, thus preventing light leaking in from the left, right and top side of the door. The gap between the lower side of the door and the floor is covered with neoprene, and a draught excluder. For the other doors in the lab the gap between the door frame and door is closed with aluminium tape, whereas the lower gap is covered by neoprene. Other light leaks are covered by aluminium tape upon inspection.\\
\begin{figure}[tbh]
\centering
	\includegraphics[width=\textwidth]{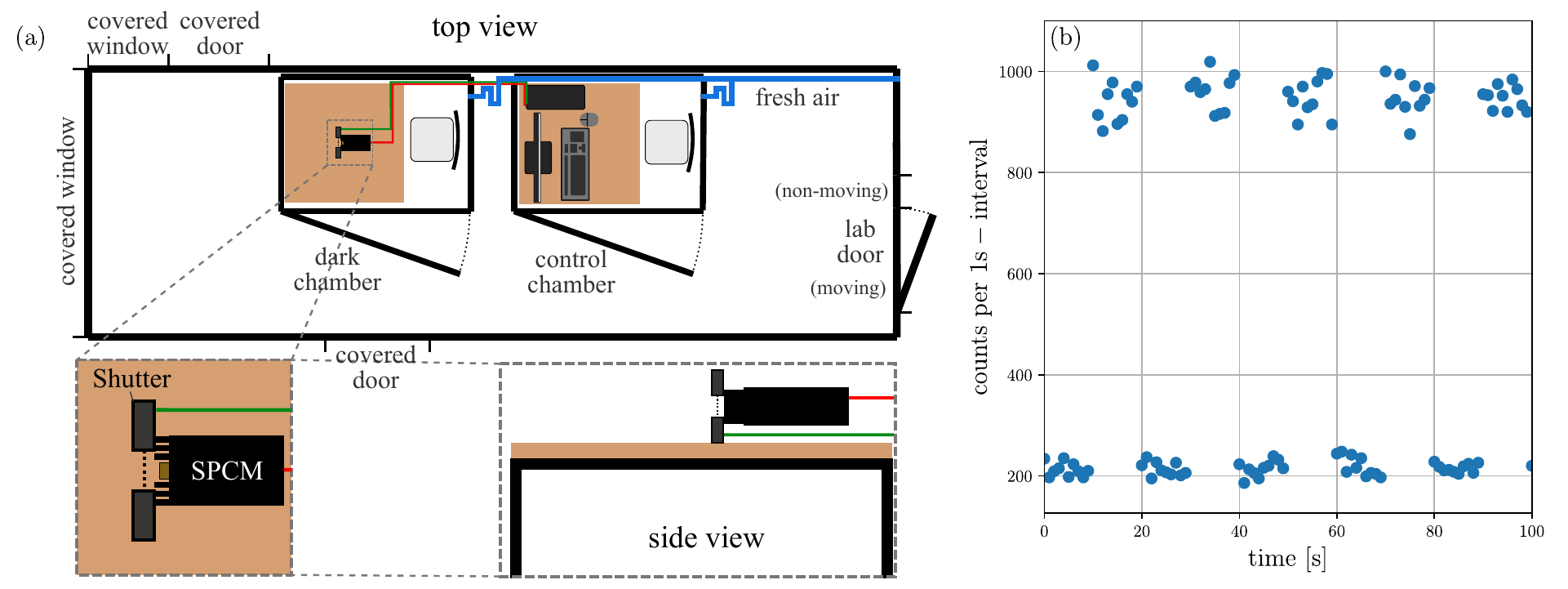}
	\caption{Schematic overview of setup and method. (a) A dark lab is created by covering all lab windows and doors using sheets of black foam board, aluminium tape and neoprene. Within the lab two chambers are placed, one control chamber and the dark chamber. A shuttered SPCM is located inside the dark chamber to assess the level of darkness. (b) Opening and closing the shutter enables us to measure the difference in counts on the SPCM between the two shutter states. This is an example of data for a light level of approximately $\SI{750}{cnt/s}$ to illustrate the method. Count fluctuations are due to detector shot noise.}\label{figLab}
\end{figure}
Within the darkened lab, we additionally place two chambers (inverted grow tents\footnote{Phantom $90\times 150\times \SI{200}{cm}$ ($W\times L\times H$)}). The white outer side reflects light impinging on the tent, whereas the black inner side absorbs photons that have entered or are created in the chamber. The first chamber, the control chamber, contains the apparatuses controlling the experiment and prevents light produced by these devices to enter the lab. The latter chamber is our dark chamber. Both chambers are connected to a ventilation system to provide fresh air. Using a bended black hose, we prevent light leaking from the control chamber into the dark chamber. A schematic overview of the lab is presented in figure~\ref{figLab}(a) and some photographic details can be found in appendix \ref{appLabPhotos}.\\

To measure light levels in the dark chamber and lab, we use a single photon counting module (SPCM\footnote{PerkinElmer SPCM-AQRH-$14$}). This SPCM is located on a desk table in the dark chamber and detects single photons ($\lambda\approx 400-\SI{1060}{nm}$) using a Silicon avalanche photodiode with a $\SI{180}{\mu m}$-diameter circular active area. The detector has a (time-varying) dark count rate of approximately $R_{\text{d}}=\SI{200}{s^{-1}}$. The SPCM is connected to a shutter\footnote{Uniblitz VS$14$ with D$122$ shutter driver} positioned approximately $\SI{35}{mm}$ in front of the photodiode using an in-house built adapter. With a shutter diameter of $\SI{14}{mm}$ this implies the effective SPCM half-opening angle equals $\arctan(0.5\cdot14/35)=11.3^{\circ}$. One side of the shutter is black, whereas the other has a metallic finish, and we mount it such that the black side is facing the SPCM. This setup allows to measure counts while the shutter is open ($C_{\text{O}}$, measuring dark counts plus light counts) or closed ($C_{\text{C}}$, measuring dark counts only).  The counts from the detector are collected by and the shutter is controlled from a data acquisition (DAQ) card\footnote{National Instruments NI-$6341$} connected to a computer in the control chamber.\\
In this work we are concerned with measuring differences in $C_{\text{O}}$ and $C_{\text{C}}$, which we evaluate using the difference distribution $\Delta\sim C_{\text{O}}-C_{\text{C}}$. To this end we count the SPCM's TTL pulses during $\SI{1}{s}$-intervals in consecutive blocks of $\SI{10}{s}$ in which the shutter is open or closed using the internal $\SI{80}{MHz}$-clock of the DAQ-card for timing. Hence, we measure for ten $\SI{1}{s}$-intervals with shutter open, ten intervals with shutter closed, ten intervals open, etc., see figure~\ref{figLab}(b). The shutter opens/closes during the first interval of the respective open/close blocks. These intervals are discarded for further analysis, in order to prevent the shutter motion influencing $C_{\text{O}}$ and $C_{\text{C}}$ in any manner. By alternating the open/close blocks, the effect of time-variations in dark count rate $R_{\text{d}}$ can be mitigated. The difference distribution $\Delta$ is obtained by subtracting the number of counts in the $i^{\text{th}}$ interval of $C_{\text{O}}$ and $C_{\text{C}}$. Although these intervals are $\SI{10}{s}$ apart in time, $R_{\text{d}}$ is not expected to vary within such a short time frame.\\

\section{Bayesian assessment of darkness}\label{secBayes}
To perform a Bayesian assessment of darkness, we have written an analysis programme in \textsc{R} \cite{Rcore2017}. We follow a Bayesian approach using \textsc{RStan} \cite{Rstan} fitting $\Delta$ measured within the dark chamber with a normal likelihood $\Delta\sim \mathcal{N}(\mu_{\Delta},\sigma_{\Delta}^2)$. Thus estimates are obtained for the unknown mean $\mu_{\Delta}$ and standard deviation $\sigma_{\Delta}$ of $\Delta$ using Markov-Chain Monte Carlo (MCMC) sampling. As such we sample from the posterior distributions of these parameters
\begin{equation}\label{eqBayes}
	p\left(\mu_{\Delta},\sigma_{\Delta}|\Delta\right)\propto p\left(\Delta|\mu_{\Delta},\sigma_{\Delta}\right)p\left(\mu_{\Delta},\sigma_{\Delta}\right).
\end{equation} 
Here, $p\left(\Delta|\mu_{\Delta},\sigma_{\Delta}\right)$ is the likelihood of the data and $p\left(\mu_{\Delta},\sigma_{\Delta}\right)$ is the prior distribution for $\mu_{\Delta}$ and $\sigma_{\Delta}$, which we will denote $p\left(\mu_{\Delta}^{(0)},\sigma_{\Delta}^{(0)}\right)$ from now on. Consequently, we will use the short hand notation $p\left(\mu_{\Delta}^{(1)},\sigma_{\Delta}^{(1)}\right)$ for the posterior distribution.  $\mu_{\Delta}$ and $\sigma_{\Delta}^2$ are assigned the priors
\begin{align}
\begin{aligned}
	\mu_{\Delta}^{(0)}&\sim\mathcal{N}\left(\mu_0,\sigma_0^2\right)\\
	\left(\sigma_{\Delta}^{2}\right)^{(0)}&\sim\Gamma\left(\alpha_0,\beta_0\right),
\end{aligned}
\end{align}
in which the prior mean $\mu_0$ and variance $\sigma_0^2$ of the normal prior, and the shape parameters $\alpha_0$ and $\beta_0$ of the gamma prior follow from the data
\begin{alignat}{2}\label{eqShapeDeltaAB}
\begin{aligned}
	\mu_0&=\max\left(0,\mean{\Delta}\right)	\qquad && \beta_0=\dfrac{\var{\Delta}\N{\Delta}}{2f_{\sigma}\var{\Delta}^2}=\dfrac{\N{\Delta}}{2f_{\sigma}\var{\Delta}}\\
	\sigma_0^2&=\dfrac{f_{\mu}\var{\Delta}}{\N{\Delta}}		\qquad && \alpha_0=\var{\Delta}\beta_{0}.
\end{aligned}
\end{alignat}
In these equations the variance of the mean (for $\sigma_0^2$, $\var{\cdot}/\N{\cdot}$, where $\N{\cdot}$ refers to the total number of measurements performed) and the variance on the variance (for $\beta_0$, $2\var{\cdot}^2/\N{\cdot}$ \cite{CasselaBerger2002}) of the normal distribution can be recognised. Both these variances are multiplied by a factor, $f_{\mu}$ and $f_{\sigma}$ respectively, to broaden the prior distributions with respect to their expected distributions. As such the priors become less informed. The choice of data-dependent prior parameters was made to make the analysis programme also applicable in situations in which we do expect to measure an unknown amount of light counts. In such a case it seems a reasonable choice to add minimal information to the prior distributions in the form of an expected value, and to decrease the impact of this information by introducing prior broadening, thereby directly allowing for a sensitivity analysis. It should be noted that in the following we choose the prior broadening factors equal, $f_{\mu}=f_{\sigma}=f$, and that for our data $\mu_0=0$ (see below).\\

\begin{figure}
\centering
	\includegraphics[width=\textwidth]{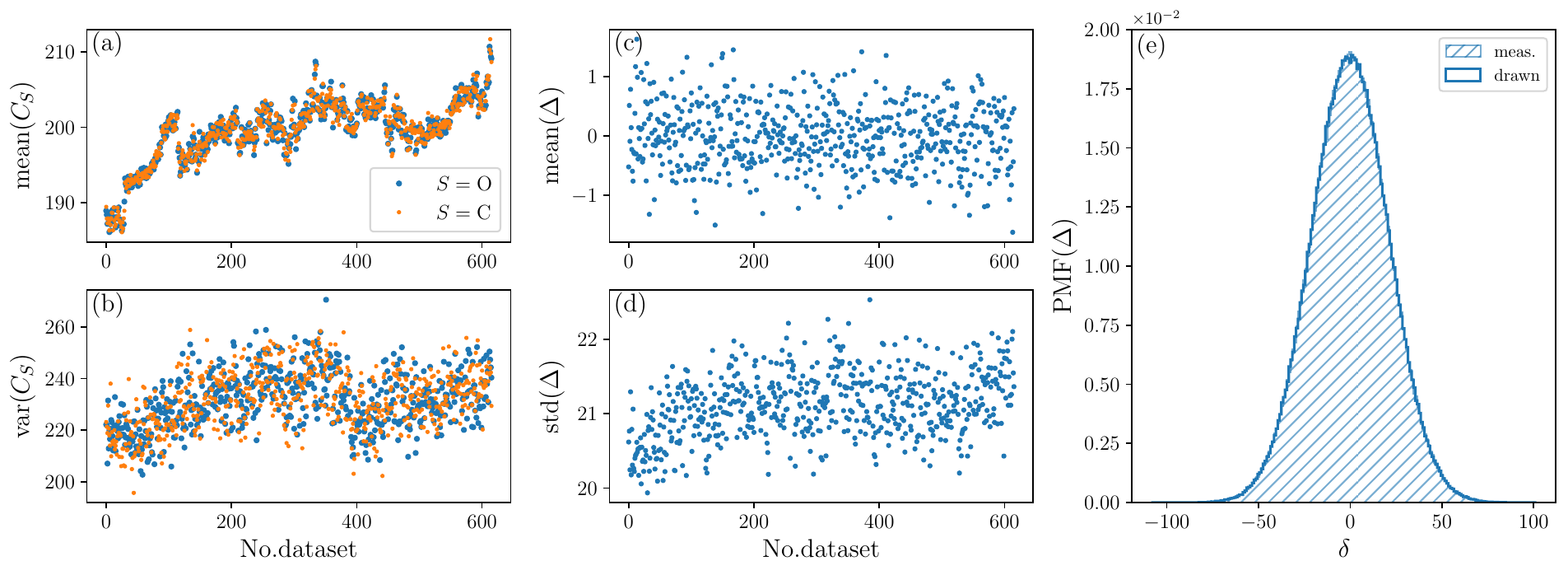}
	\caption{Measured results for distributions $C_{\text{O}}$, $C_{\text{C}}$ and $\Delta$ after $\SI{616}{h}$ of measurement. (a)-(b) Mean and variance of distributions $C_{\text{O}}$ and $C_{\text{C}}$ per $\SI{1}{h}$-dataset. These values vary as a result of fluctuations in $R_{\text{d}}$. (c)-(d) Mean and standard deviation of $\Delta$ per $\SI{1}{h}$-dataset. These values are found to be constant. (e) Distribution of $\Delta$. The hatched measured PMF is overlain with a PMF of samples drawn from a normal distribution with the same mean and standard deviation as the data.}\label{figCDeltaMeas}
\end{figure}

We assess the darkness in our dark chamber based on $\SI{616}{h}$ of measurements. This yields $0.9\cdot 1800\cdot 616=997920$ $\SI{1}{s}$-intervals sampling the distribution $\Delta$.\\
The measurements are summarised in figure \ref{figCDeltaMeas}. Here, in figure~\ref{figCDeltaMeas}(a), it is observed that the distributions $C_{\text{O}}$ and $C_{\text{C}}$ shift over time as a result of fluctuations in $R_{\text{d}}$. As a result, these distributions are not directly comparable to assess the level of darkness in the dark chamber. This issue is no longer observed for $\Delta$ for which mean and standard deviation have become constant in time, see figure~\ref{figCDeltaMeas}(b). The measured distribution $\Delta$ has been depicted in figure~\ref{figCDeltaMeas}(c). The measured point mass function (PMF) is overlain with a PMF drawn from Gaussian distribution with the same mean ($\SI{-4.14e-3}{cnt/s}$) and standard deviation ($\SI{21.1}{cnt/s}$), showing excellent agreement.\\

\begin{figure}
\centering
	\includegraphics[width=\textwidth]{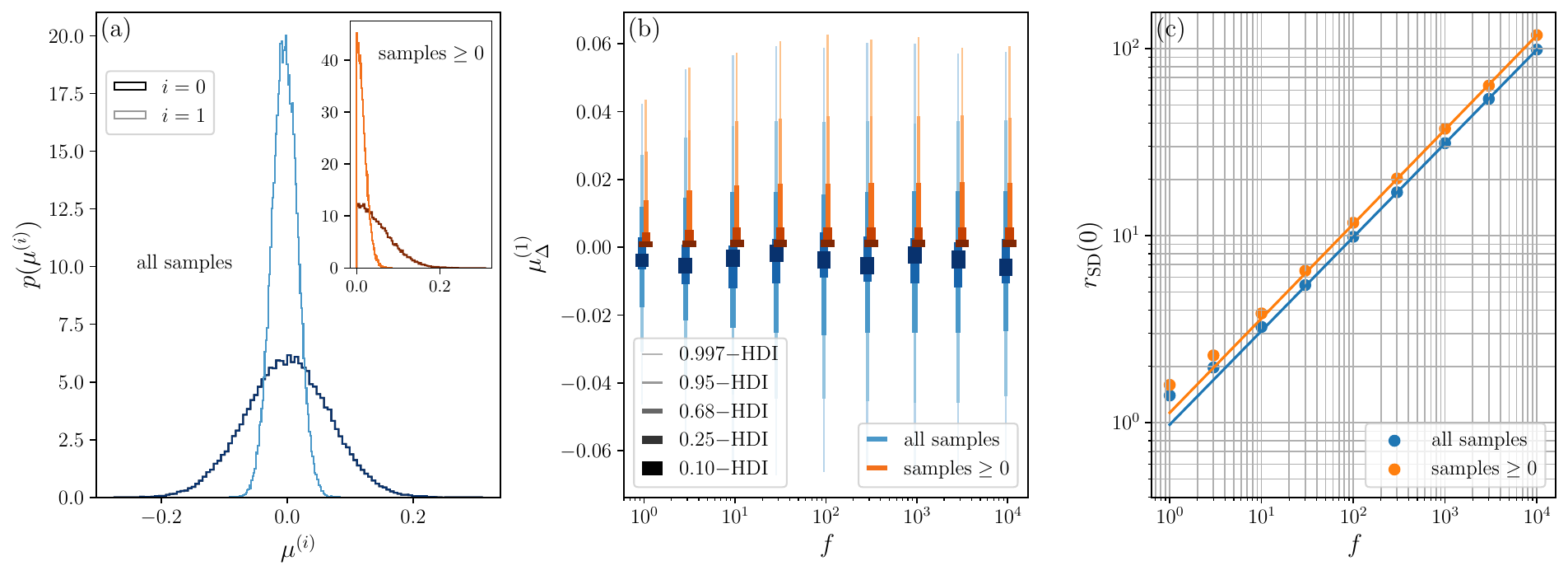}
	\caption{\textsc{Rstan} fit results for $\mu_{\Delta}$. (a) Prior and posterior distribution for $f=10$. These distributions are analysed in terms of the probability of positive direction, HDIs and Savage-Dickey ratios. The inset shows the positive part of the distributions, stemming from the requirement that the number of photons still present must be strictly positive. (b) Posterior distribution as a function of $f$ visualised using indicated HDIs. For $f\geq 10$, the posterior is constant, indicating $f=10$ is the minimum value for obtaining a posterior independent of the prior. (c) Savage-Dickey ratio for null result (no light present) as function of $f$. Data points for $f\geq 10$ are fitted using a power-function.}\label{figBayesResults}
\end{figure}

In our analysis of the data, we focus on the parameter $\mu_{\Delta}^{(1)}$ and quantify evidence for $\mu_{\Delta}^{(1)}=0$ and $\mu_{\Delta}^{(1)}\neq 0$. The analysis results for $\mu_{\Delta}^{(1)}$ are shown in figure \ref{figBayesResults}. In figure~\ref{figBayesResults}(a) the prior and posterior distribution for $\mu_{\Delta}$ are depicted for a prior broadening factor of $f=10$. These density plots are based on  $\SI{6e4}{MCMC\ samples}$ drawn from $p(\mu_{\Delta}^{(1)})$ in \textsc{RStan}. The inset of this figure features the positive part of the distributions, corresponding to the requirement that the number of photons still present in the dark chamber must be strictly positive. From the full distribution we obtain the probability of positive direction (adapted from the probability of direction \cite{Makowskietal2019} for purely positive effects), $\text{pd}^{+}=p(\mu_{\Delta}^{(1)}>0)$, as indication of whether light has been detected ($\text{pd}^{+}>0.5$), and the Savage-Dickey (SD) ratio \cite{Dickey1971} as a Bayes' factor for absolute darkness. This ratio is given by $r_{\text{SD}}(0)=p(\mu_{\Delta}^{(1)}=0)/p(\mu_{\Delta}^{(0)}=0)$, where $p(\mu_{\Delta}^{(1)}=0)$ is obtained from a logspline-fit of the MCMC samples' density using the \textsc{logspline}-library available for \textsc{R} \cite{Rlogspline}. $p(\mu_{\Delta}^{(0)}=0)$, on the other hand, follows directly from the prior distribution. We perform two hypothesis tests using the SD-ratio:
\begin{itemize}
	\item $H_0: \mu_{\Delta}=0$ vs. $H_1:\mu_{\Delta}\neq 0$ (using full distributions -- evidence for $\mu_{\Delta}$ not different from $0$ given $\mu_{\Delta}$ can be positive or negative), and
	\item $H_0: \mu_{\Delta}=0$ vs. $H_1:\mu_{\Delta}\geq 0$ (using positive part of distributions -- evidence for $\mu_{\Delta}$ not different from $0$ given $\mu_{\Delta}$ is strictly positive).
\end{itemize} 
Finally, from the positive part of the distribution highest density intervals (HDIs) are obtained as an upper limit to the light-counts detected.\\ 
In figure~\ref{figBayesResults}(b) and (c) the results of these measures can be observed for fits with different prior broadening factors $f$. Figure (b) depicts the posterior distribution for $\mu_{\Delta}$ using indicated HDIs for both the full distributions as well as its positive part. As can be seen, the distributions become constant for $f\geq 10$, indicating that for lower $f$ the fit is influenced by the choice of prior, which leads to an underestimation of HDIs. Therefore we will not consider these fits in the following. From the full distribution we find $\mu_{\Delta}^{(1)}=(-4.19\pm 21.0)\times \SI{10^{-3}}{ cnt/s}$ (mean $\pm$ standard deviation, $0.95$-HDI $[-4.50,3.71]\times \SI{10^{-2}}{cnt/s}$) and we obtain a $\text{pd}^{+}$ of $0.42$, indicating uncertainty in the direction of $\mu_{\Delta}^{(1)}$ \cite{Makowskietal2019}. Taking the $0.95$-HDI as an upper limit to the number of light-counts detected, the positive part of the posterior yields the number of light-counts $<\SI{0.039}{cnt/s}$. On the other hand, as observed in figure~\ref{figBayesResults}(c), the SD-ratio depends on $f$ by a sqrt-dependence, as follows from a fit of the form $r_{\text{SD}}(0)=af^b$. Based on these results, the evidence for absolute darkness is substantial ($r_{\text{SD}}(0)>10^{1/2}$), strong ($r_{\text{SD}}(0)>10^{1}$), very strong ($r_{\text{SD}}(0)>10^{3/2}$) or even decisive ($r_{\text{SD}}(0)>10^{2}$) \cite{Jeffreys1961} depending on the choice of $f$.

To assess the robustness of our results, we performed an alternative analysis using a Cauchy prior as proposed by, e.g., Refs. \cite{GronauLyWagenmakers2020,Jeffreys1961,Rouderetal2009}. The advantage of such a prior is that the Bayes' factor can be calculated analytically. We use a location parameter of $0$ and a scale parameter of $0.707$ which is the default prior in the popular Bayesian analysis software package JASP \cite{Loveetal2019}. This choice is comparable to setting $f=0.707\N{\Delta}=705529$ in our own analysis programme. Results indicate that the Bayes' factor supporting the hypothesis that $\mu_{\Delta}=0$ (given $\mu_{\Delta}\geq 0$) equals $1029.59$, which is in agreement with $r_{\text{SD}}(0)\approx 1011$ given by our analysis. This indicates that there is decisive evidence in the data supporting the hypothesis that $\mu_{\Delta}=0$.

\section{Environment-Attenuation approach}\label{secBright}
Additionally, we assess the level of darkness in our dark chamber using the EA approach. In this approach we measure the light level in the lab, $\bar{\Delta}_{\text{lab}}$, as well as the attenuation of the dark chamber, $A_{\text{c}}$, using a bright light source. As such, an estimate of the light level in the dark chamber can be obtained as $\bar{\Delta}_{\text{EA}}\approx \bar{\Delta}_{\text{lab}}/A_{\text{c}}$.\\ 
For our EA measurements the SPCM is rotated by $90^{\circ}$ with respect to figure \ref{figLab}(a), such that it can detect light coming from the lab. In this position the darkness level of the lab is assessed, while the dark chamber is open and found to be $\bar{\Delta}_{\text{lab}}=1.82\pm \SI{0.03}{cnt/s}$ during $\SI{246}{h}$ of measurements ($398520$ samples).\\ 
Secondly we estimate $A_{\text{c}}$. To this end, a bright $\SI{530}{nm}$-light emitting diode\footnote{Lumileds LXZ$1$-PM$01$} (LED) is mounted outside of the (still open) dark chamber pointing towards the SPCM. At a current of $\SI{800}{mA}$ the light level of this LED was measured to be $10^{9.10\pm 0.01}\mathrm{ cnt/s}$ as detailed in appendix \ref{appEA}. Thereafter the light level of the bright LED on the SPCM is measured while the dark chamber is closed. This level was found to be $0.47\pm \SI{0.09}{cnt/s}$, based on $\SI{36}{h}$ of measurement ($58320$ samples). As such, we find that $A_{\text{c}}\approx 10^{9.10\pm 0.01}/(0.47\pm 0.09) = (2.8\pm 0.6)\times 10^9$.\\
From $\bar{\Delta}_{\text{lab}}$ and $A_{\text{c}}$, we expect the light level in the dark chamber to be $\Delta_{\text{EA}}\approx 1.82\pm 0.03/[(2.8\pm 0.6)\times 10^9]=(6.9\pm 1.8)\times\SI{10^{-10}}{cnt/s}$.

\section{Discussion}\label{secDiscussion}
The EA approach gives an estimate of the level of darkness orders of magnitude lower than the direct method ($(6.9\pm 1.8)\times\SI{10^{-10}}{cnt/s}$ (EA) vs. $<\SI{0.039}{cnt/s}$ (direct)). This discrepancy is inherent to the methods: in the EA approach a (low) light level in the lab is divided by the (large) attenuation of the dark chamber, whereas the direct method relies on reducing the standard deviation on $\mu_{\Delta}$. However, as mentioned before, difficulties in interpretation arise for the former method, most importantly because it is insensitive for light generated inside the dark chamber itself. To clarify: we measured a darkness level of $0.47\pm\SI{0.09}{cnt/s}$ in the closed dark chamber, while the bright LED outside the chamber was switched on (see section \ref{secBright}). However, within the EA approach one cannot know whether \emph{all} detected light originates from the bright LED. The light might as well be (partly) emitted from within the chamber itself by, e.g., phosphorescence. This problem of interpretation is circumvented by the direct method.\\ 
Yet, a similar argument can be made for the direct method. As shown in appendix \ref{appMetallic} we infer a likely negative value of $\mu_{\Delta}^{(1)}=(-4.48\pm 2.31)\times\SI{10^{-2}}{cnt/s}$ ($0.95$-HDI $[-\SI{9.0e-2}{},\SI{6.0e-4}{}]\mathrm{\ cnt/s}$) in case the other (metallic-finished) side of the shutter faces the detector. We attribute this result to the reflection of SPCM-emitted flash photons from the shutter. This implies that the results shown in section \ref{secBayes} might be offset by an unknown amount, e.g., if the detector would measure $\SI{10.000}{cnt/s}$ when the shutter is open due to background light in the dark chamber, and it would detect $\SI{10.004}{cnt/s}$ while the shutter is closed as a result of flash photons, we would obtain the same result. In appendix \ref{appFlashReflection} we provide a simple model taking the contribution of reflected flash photons into account and find that the mentioned offset is absent if the black side of the shutter faces the detector.\\

It should be noted that the Bayesian approach is a slow method. After hundreds of hours of measurement, the upper limit on the light-clicks detected is still in the order of $10^{-2}$. This is a direct result of the variance in the rate of dark counts produced in the SPCM in combination with the $\sqrt{N}$-law governing the width of the posterior distribution for $\mu_{\Delta}$. In absolute darkness the $0.95$-HDI length for the upper limit of the level of darkness will be given by
\begin{equation}
	L_{0.95}^{\text{dark}}=1.96\sqrt{\dfrac{2\var{R_{\text{d}}}}{\N{\Delta}}},
\end{equation}
where the factor $1.96$ is the Gaussian $z$-score for the $0.95$-HDI, and the variance in dark count rate is multiplied by $2$, because we are subtracting two distributions to obtain $\Delta$, $C_{\text{C}}$ from $C_{\text{O}}$. From this equation it is evident that using a detector with smaller $\var{R_{\text{d}}}$, such as a superconducting nanowire single photon detector (SNSPD) \cite{EsmaeilZadehetal2021}, is more beneficial for reaching a lower upper limit for the number of light-counts. For such detectors dark count rates of $10^{-4}\,\SI{}{s^{-1}}$ have been reported. Assuming $R_{\text{d}}$ of these detectors to be of the same order, our result could be reproduced with only $\SI{0.5}{s}$ of measurement time. The disadvantage of SNSPDs, however, is their operation at cryogenic temperatures, which greatly increases the experimental complexity. Still, it should be noted that even with such devices an upper limit of $0$ will never be achieved. Alternatively, single-photon sensitive cameras, such as an electron multiplying charge-coupled device (EMCCD) \cite{Lantzetal2008}, could be utilised to study positional variation in background light, such that remaining light sources can be discovered and mitigated. This method also comes with the cost of increased experimental complexity, however.

Still, since the probability of positive direction $\text{pd}^{+}=0.42$ is uncertain as of which direction $p(\mu_{\Delta}^{(1)})$ favours, and $r_{\text{SD}}(0)$ yields substantial to decisive evidence\footnote{Note that $r_{\text{SD}}(\mu_{\delta})$ would also give evidence for $\mu_{\delta}>0$, but for our data the result peaks for $\mu_{\delta}=0$, see figure \ref{figBayesResults}(a).} for $\mu_{\Delta}^{(1)}=0$, our results suggests the dark chamber to be absolutely dark. The amount of evidence for absolute darkness depends critically on the prior broadening factors $f_{\mu}$ and $f_{\sigma}$. In analyzing the data of this and similar experiments, the choice for these factors could be based on prior knowledge, e.g., from previous experiments. If such knowledge is absent (as in the present study), we recommend using $f = 0.707\N{\Delta}$ so that the resulting (uninformative) prior is comparable to the default prior in JASP. However, it is recommendable to conduct a sensitivity analysis, similar to our analysis in figure \ref{figBayesResults}c in order to study the robustness of the result.\\
In terms of the human eye these results can be understood as follows. The shutter diameter and position with respect to the SPCM photodiode are such that the half-opening angle is approximately equal to the half-opening angle of the retina due to the pupil (a pupil diameter equal to $\SI{10}{mm}$ and pupil-retina distance of $\SI{24}{mm}$ yields a half-opening angle of $11.7^{\circ}$, only slightly larger than the $11.3^{\circ}$ for the SPCM). Neglecting the quantum efficiency of the SPCM, the loss in the vitreous body and the sensitivity of the retina's rods and cones, our results compare directly to the number of $400-\SI{1060}{nm}$-background photons reaching an area of $\pi\cdot(\SI{180}{\mu m}/2)^2$ of the retina. Hence, if during actual psychophysical experiments light pulses reach the retina within a circular area of diameter $d$, the upper limit to the number of counts caused by background light within this area is $0.039\cdot d/(\SI{180}{\mu m})\,\SI{}{s^{-1}}$. Thus, if the light pulses enter the eye under a Maxwellian view with half-opening angle of $0.4^{\circ}$ ($d=\SI{0.34}{mm}$, see section \ref{secIntroduction}) this implies that the upper limit of background photons is given by $\SI{0.074}{cnt/s}$.

\section{Conclusions}\label{secConclusions}
We have assessed the level of darkness in a dark chamber for psychophysical experiments using a single-photon avalanche photodiode. By placing the detector behind a shutter, we could compare the detector shutter-open and shutter-closed count distributions, thus providing a direct measurement of the level of darkness. From a Bayesian fit of the data taken over $\SI{616}{h}$ we found the light level to be below $\SI{0.039}{cnt/s}$ ($0.95$-HDI), while the posterior distribution exhibited a probability of positive direction of $0.42$ (uncertain, whether an effect is present) and substantial to decisive evidence (based on the Savage-Dickey ratio) for absolute darkness depending on the prior distribution chosen. On the other hand we estimated a light level of $(6.9\pm 1.8)\times\SI{10^{-10}}{cnt/s}$ from an environment-attenuation approach, in which the light level in the lab and the attenuation of the dark chamber were evaluated. The exact interpretation of this last number, however, is problematic. Although we have conducted this work in light of psychophysical experiments, the method is directly applicable to all experiments in which a null-background is of high-importance.

\subsection*{Acknowledgements}
We acknowledge funding from NWO/OCW (Quantum Software Consortium, No. $024.003.037$), from the Dutch
Ministry of Economic Affairs (Quantum Delta NL), and from the European Union’s Horizon $2020$ research and innovation program under Grant Agreement No. $862035$ (QLUSTER).

\subsection*{Data availability}
The data underlying this manuscript and example evaluation scripts are available in Ref. \cite{Data}.

\bibliographystyle{unsrt}
\bibliography{darkness_arxiv_v2.bbl}

\begin{thebibliography}{10}

\bibitem{KoenigHofer2011}
D.~Koenig and H.~Hofer.
\newblock {The absolute threshold of cone vision}.
\newblock {\em J. Vis.}, 11:21--21, 2011.

\bibitem{Hechtetal1941}
S.~Hecht, S.~Schlaer, and M.~H. Pirenne.
\newblock Energy at the threshold of vision.
\newblock {\em Science}, 93:585--587, 1941.

\bibitem{Sakitt1972}
B.~Sakitt.
\newblock Counting every quantum.
\newblock {\em J. Physiol.}, 223:131--150, 1972.

\bibitem{Teichetal1982}
M.~C. Teich, P.~R. Prucnal, G.~Vannucci, M.~E. Breton, and W.~J. McGill.
\newblock Multiplication noise in the human visual system at threshold: 1.
  quantum fluctuations and minimum detectable energy.
\newblock {\em J. Opt. Soc. Am.}, 72:419--431, 1982.

\bibitem{Tinsleyetal2016}
J.~N. Tinsley, M.~I. Molodtsov, R.~Prevedel, D.~Wartmann, J.~Espigul{\'e}-Pons,
  M.~Lauwers, and A.~Vaziri.
\newblock Direct detection of a single photon by humans.
\newblock {\em Nature Commun.}, 7:12172, 2016.

\bibitem{Reepetal2023}
T.~H.~A. van~der Reep, D.~Molenaar, W.~L\"{o}ffler, and Y.~Pinto.
\newblock Quantum detector tomography applied to the human visual system: a
  feasibility study.
\newblock {\em J. Opt. Soc. Am. A}, 40:285--293, 2023.

\bibitem{Fieldetal2019}
G.~D. Field, V.~Uzzell, E.~J. Chichilnisky, and F.~Rieke.
\newblock Temporal resolution of single-photon responses in primate rod
  photoreceptors and limits imposed by cellular noise.
\newblock {\em J. Neurophysiol.}, 121:255--268, 2019.
\newblock PMID: 30485153.

\bibitem{Westheimer1966}
G.~Westheimer.
\newblock The maxwellian view.
\newblock {\em Vision Res.}, 6:669--682, 1966.

\bibitem{Cursioetal1990}
C.~Curcio, K.~Sloan, R.~Kalina, and A.~Hendrickson.
\newblock Human photoreceptor topography.
\newblock {\em J. Comp. Neurol.}, 292:497--523, 1990.

\bibitem{DentonPirenne1954}
E.~J. Denton and M.~H. Pirenne.
\newblock The absolute sensitivity and functional stability of the human eye.
\newblock {\em J. Physiol.}, 123:417--442, 1954.

\bibitem{HalettMarriottRodger1962}
P.~E. Hallett, F.~H.~C. Marriott, and F.~C. Rodger.
\newblock The relationship of visual threshold to retinal position and area.
\newblock {\em J. Physiol.}, 160:364--373, 1962.

\bibitem{Holmesetal2017}
R.~Holmes, M.~Victora, R.~F. Wang, and P.~G. Kwiat.
\newblock Measuring temporal summation in visual detection with a single-photon
  source.
\newblock {\em Vision Res.}, 140:33--43, 2017.

\bibitem{Deyetal2021}
A.~Dey, A.~J. Zele, B.~Feigl, and P.~Adhikari.
\newblock Threshold vision under full-field stimulation: Revisiting the minimum
  number of quanta necessary to evoke a visual sensation.
\newblock {\em Vision Res.}, 180:1--10, 2021.

\bibitem{Wagenmakersetal2008}
E.-J. Wagenmakers, M.~D. Lee, T.~Lodewyckx, and G.~Iverson.
\newblock Bayesian versus frequentist inference.
\newblock In H.~Hoijtink, I.~Klugkist, and P.~A. Boelen, editors, {\em Bayesian
  evaluation of informative hypotheses in psychology}, pages 181--207.
  Springer, 2008.

\bibitem{Rcore2017}
{R Core Team}.
\newblock {\em R: A Language and Environment for Statistical Computing}.
\newblock R Foundation for Statistical Computing, Vienna, Austria, 2017.

\bibitem{Rstan}
{Stan Development Team}.
\newblock {RStan}: the {R} interface to {Stan}, 2023.
\newblock R package version 2.32.5.

\bibitem{CasselaBerger2002}
G.~Cassela and R.~L. Berger.
\newblock {\em Statistical Inference}.
\newblock Duxbury Press, Pacific Grove, USA, $2^{\text{nd}}$ edition, 2002.

\bibitem{Makowskietal2019}
D.~Makowski, M.~S. Ben-Shachar, and D.~L\"udecke.
\newblock {bayestestR: Describing Effects and their Uncertainty, Existence and
  Significance within the Bayesian Framework}.
\newblock {\em J. Open Source Softw.}, 4:1541, 2019.

\bibitem{Dickey1971}
J.~M. Dickey.
\newblock {The weighted likelihood ratio, linear hypotheses on normal location
  parameters}.
\newblock {\em Ann. Math. Statist.}, 42:204--223, 1971.

\bibitem{Rlogspline}
C.~Kooperberg.
\newblock {\em logspline: Routines for Logspline Density Estimation}, 2024.
\newblock R package version 2.1.21.

\bibitem{Jeffreys1961}
H.~Jeffreys.
\newblock {\em Theory of Probability}.
\newblock Oxford University Press, Oxford, UK, $3^{\text{rd}}$ edition, 1961.

\bibitem{GronauLyWagenmakers2020}
Q.~F. Gronau, A.~Ly, and E.-J. Wagenmakers.
\newblock Informed bayesian t-tests.
\newblock {\em Am. Stat.}, 74:137--143, 2020.

\bibitem{Rouderetal2009}
J.~N. Rouder, P.~L. Speckman, D.~Sun, R.~D. Morey, and G.~Iverson.
\newblock Bayesian $t$ tests for accepting and rejecting the null hypothesis.
\newblock {\em Psychon. Bull. Rev.}, 16:225–--237, 2009.

\bibitem{Loveetal2019}
J.~Love, R.~Selker, M.~Marsman, T.~Jamil, D.~Dropmann, J.~Verhagen, A.~Ly,
  Q.~F. Gronau, M.~\v{S}m\'ira, S.~Epskamp, D.~Matzke, A.~Wild, P.~Knight,
  J.~N. Rouder, R.~D. Morey, and E.-J. Wagenmakers.
\newblock {JASP}: Graphical statistical software for common statistical
  designs.
\newblock {\em J. Stat. Softw.}, 88:1–--17, 2019.

\bibitem{EsmaeilZadehetal2021}
I.~Esmaeil~Zadeh, J.~Chang, J.~W.~N. Los, S.~Gyger, A.~W. Elshaari,
  S.~Steinhauer, S.~N. Dorenbos, and V.~Zwiller.
\newblock Superconducting nanowire single-photon detectors: A perspective on
  evolution, state-of-the-art, future developments, and applications.
\newblock {\em App. Phys. Lett.}, 118:190502, 2021.

\bibitem{Lantzetal2008}
E.~Lantz, J.-L. Blanchet, L.~Furfaro, and F.~Devaux.
\newblock {Multi-imaging and Bayesian estimation for photon counting with
  EMCCDs}.
\newblock {\em Mon. Not. R. Astron. Soc.}, 386:2262--2270, 2008.

\bibitem{Data}
T.~van~der Reep, D.~Molenaar, and W.~L\"offler.
\newblock Data: Measurement of darkness (1.0.0).
\newblock DOI: 10.5281/zenodo.13951375, 2024.

\bibitem{Lacaitaetal1993}
A.~L. Lacaita, F.~Zappa, S.~Bigliardi, and M.~Manfredi.
\newblock On the bremsstrahlung origin of hot-carrier-induced photons in
  silicon devices.
\newblock {\em IEEE Trans. Electron Devices}, 40:577--582, 1993.

\bibitem{Kurtsieferetal2001}
C.~Kurtsiefer, P.~Zarda, S.~Mayer, and H.~Weinfurter.
\newblock The breakdown flash of silicon avalanche photodiodes-back door for
  eavesdropper attacks?
\newblock {\em J. Mod. Opt.}, 48:2039–--2047, 2001.

\bibitem{Wareetal2007}
M.~Ware, A.~Migdall, Bienfang.~J. C., and S.~V. Polyakov.
\newblock Calibrating photon-counting detectors to high accuracy: background
  and deadtime issues.
\newblock {\em J. Mod. Opt.}, 54:361--372, 2007.

\end{thebibliography}
\vspace{0.5cm}

\newpage
\noindent{\Large{\textbf{Appendices}}} \\[-0.7cm]
\appendix
\renewcommand{\thefigure}{A\arabic{figure}}
\renewcommand{\thesection}{A\arabic{section}}
\renewcommand{\thesubsection}{A\arabic{section}.\arabic{subsection}}
\renewcommand{\theequation}{A\arabic{equation}}
\renewcommand{\thetable}{A\arabic{table}}

\section{Photographic details of the lab}\label{appLabPhotos}
\begin{figure}[b!]
\centering
	\includegraphics[width=\textwidth]{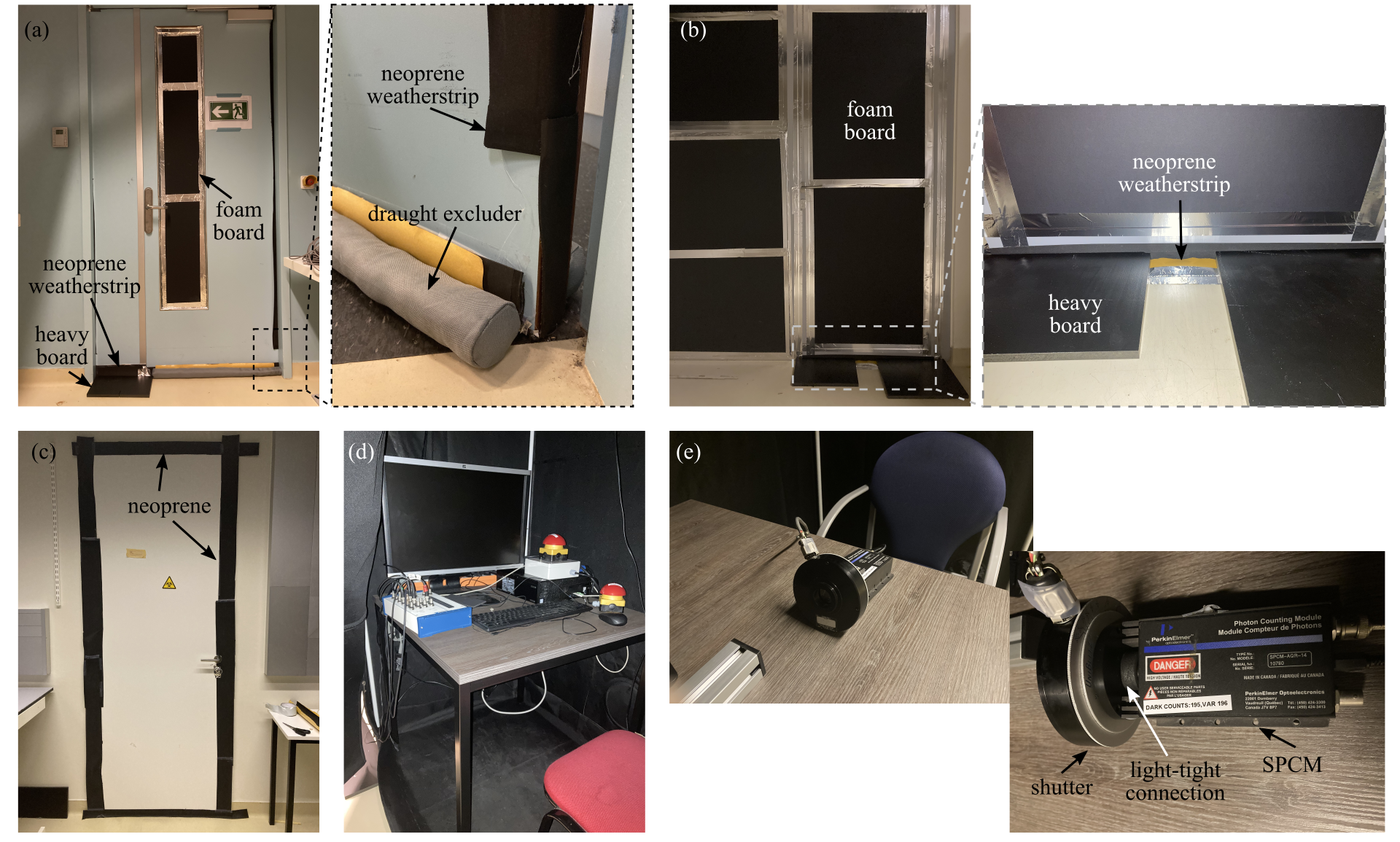}
	\caption{Photographic details of our lab. (a) The lab door, (b) the covered window and door at the top side of figure~\ref{figLab}(a), (c) the covered door at the bottom side of figure~\ref{figLab}(a), (d) the control chamber and (e) the SPCM connected to the shutter.}\label{figDarklab-photos}
\end{figure}

In this appendix we show photographic details of the measures we have taken to darken the lab, see figure~\ref{figDarklab-photos}. Figure~\ref{figDarklab-photos}(a) shows the lab door, figure ~\ref{figDarklab-photos}(b) shows the covered window and covered door at the top of the schematic overview of the lab in figure~\ref{figLab}(a), figure~\ref{figDarklab-photos}(c) shows the covered door at the bottom of the schematic overview of the lab in figure~\ref{figLab}(a), figure~\ref{figDarklab-photos}(d) shows the inside of the control chamber, and figure~\ref{figDarklab-photos}(e) shows the setup of SPCM and shutter sitting on the desk table within the dark chamber.

\section{Bright-light calibration}\label{appEA}
The light level from the bright LED light used in our EA approach is calibrated using a tomography method. In this method, combinations of neutral density filters attenuate the LED light to levels measurable by the SPCM as $\bar{\Delta}_{\downarrow\LEDsym}^{\text{O}}=\bar{\Delta}_{\LEDsym}^{\text{O}}/\Pi_i A_{i}$. Here, $\bar{\Delta}_{(\downarrow)\LEDsym}^{\text{O}}$ is the light level of the (attenuated) LED measured while the dark chamber is open and $A_{i}$ is the attenuation of filter $i$. Taking the base-$10$ logarithm of these equations, we can find the least-squares solution of the system $\vec{\bar{\Delta}}_{\downarrow\LEDsym}^{\text{O(log)}}=\mathbf{A}\vec{x}^{\text{(log)}}$, where $\vec{x}^{\text{(log)}}=[\log_{10}(\bar{\Delta}_{\LEDsym}^{\text{O}}),\text{OD}_1,\dots, \text{OD}_n]^T$ contains the log-light level of the bright LED and the optical densities (ODs) of the filters. Matrix $\mathbf{A}$ has rows with first index $1$ (the LED is on) and indices equal to $-1$ for the filters present. The least-squares solution to this system is
\begin{equation}\label{eqLS}
	\vec{x}_{\text{LS}}^{\text{(log)}}=\left(\mathbf{A}^T\mathbf{A}\right)^{-1}\mathbf{A}^T\vec{\bar{\Delta}}_{\downarrow\LEDsym}^{\text{O(log)}}.
\end{equation}
In our measurements $5$ filters are used and we sample $\Delta_{\downarrow\LEDsym}^{\text{O}}$ for all single and double filter combinations using the method described in section \ref{secMethod} during $\SI{15}{min}$ ($405$ samples). In the analysis, we leave out measured $\Delta_{\downarrow\LEDsym}^{\text{O}}$s with a mean value in excess of $10^6$, for which the SPCM is starting to show saturation effects. The results of these measurements and the least-squares solution is presented in table \ref{tabBrightResults}, from which we find $\bar{\Delta}_{\LEDsym}^{\text{O}}=10^{9.10\pm 0.01}\mathrm{ cnt/s}$.\\ 

\begin{table}[t!]
	\centering
	\caption{Results of the environment-attenuation approach. Under `combination' the presence and absence of the bright LED and filters are indicated in the respective columns, thus amounting to the matrix $\mathbf{A}$. The column $\Delta_{\downarrow\LEDsym}^{\text{O(log)}}$ contains the measurement results giving the vector $\vec{\bar{\Delta}}_{\downarrow\LEDsym}^{\text{O(log)}}$. The stated standard deviations of the measured distribution are not taken into account into the least-squares solution presented under $\vec{x}_{\text{LS}}^{\text{(log)}}$. The solution error is estimated from the least-squares fits in which one of the filters was left out.}\label{tabBrightResults}
	\begin{tabularx}{0.71\textwidth}{|c|c|c|c|c|c|c|X|}
	\hline
		& $\Delta_{\LEDsym}^{\text{O}}$&$A_1$&$A_2$&$A_3$&$A_4$&$A_5$&$\Delta_{\downarrow\LEDsym}^{\text{O(log)}}$\\
		\hline\hline
	combination& $1$& $-1$& $0$& $0$& $0$& $0$&$5.923\pm 0.002$\\
	& $1$& $0$& $-1$& $0$& $0$& $0$&$4.886\pm 0.004$\\
	& $1$& $0$& $0$& $0$& $-1$& $0$&$5.910\pm 0.003$\\
 &$1$& $-1$& $-1$& $0$& $0$& $0$&$1.72\pm 0.16$\\
 &$1$& $0$& $-1$& $-1$& $0$& $0$&$2.02\pm 0.09$\\
 &$1$& $0$& $-1$& $0$& $-1$& $0$&$1.70\pm 0.16$\\
 &$1$& $-1$& $0$& $-1$& $0$& $0$&$3.05\pm 0.02$\\
 &$1$& $0$& $0$& $-1$& $-1$& $0$&$3.04\pm 0.02$\\
 &$1$& $-1$& $0$& $0$& $-1$& $0$&$2.73\pm 0.03$\\
 &$1$& $-1$& $0$& $0$& $0$& $-1$&$2.87\pm 0.02$\\
 &$1$& $0$& $-1$& $0$& $0$& $-1$&$1.83\pm 0.13$\\
 &$1$& $0$& $0$& $-1$& $0$& $-1$&$3.18\pm 0.01$\\
 &$1$& $0$& $0$& $0$& $-1$& $-1$&$2.85\pm 0.02$\\
 \hline\hline
 $\vec{x}_{\text{LS}}^{\text{(log)}}$&$\SI{}{cnt/s}$&OD&OD&OD&OD&OD& \\
 ($\pm 0.01$)&$10^{9.10}$& $3.18$& $4.21$& $2.87$& $3.19$& $3.06$& \\
 \hline
	\end{tabularx}
\end{table}

\section{Negative count rate for metallic shutter finish}\label{appMetallic}
\begin{figure}[b!]
\vspace{-1mm}
\centering
	\includegraphics[width=\textwidth]{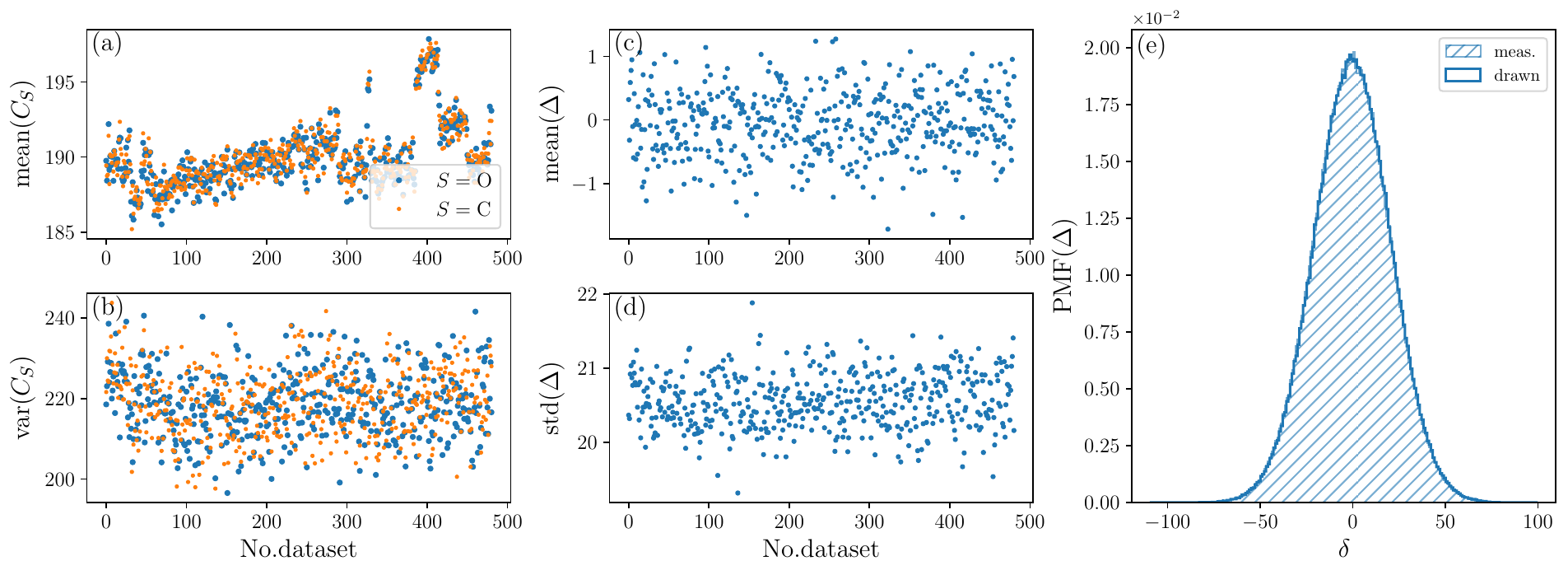}
	\caption{Measured results for distributions $C_{\text{O}}$, $C_{\text{C}}$ and $\Delta$ after $\SI{481}{h}$ of measurement. (a)-(b) Mean and variance of distributions $C_{\text{O}}$ and $C_{\text{C}}$ per $\SI{1}{h}$-dataset. (c)-(d) Mean and standard deviation of $\Delta$ per $\SI{1}{h}$-dataset. (e) Distribution of $\Delta$. The hatched measured PMF is overlain with a PMF of samples drawn from a normal distribution with the same mean and standard deviation as the data.}\label{figCDeltaMeas_v2}
\end{figure}

\begin{figure}[tbh]
\centering
	\includegraphics[width=\textwidth]{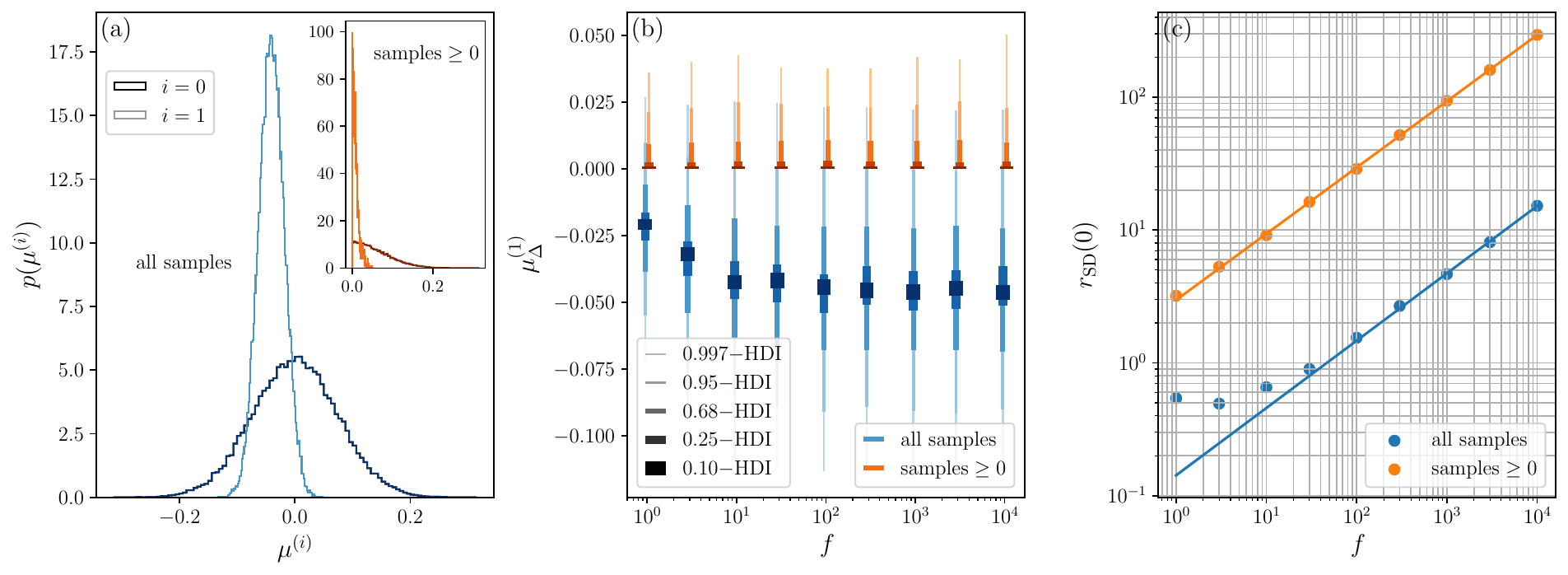}
	\caption{\textsc{Rstan} fit results for $\mu_{\Delta}$ for the dataset in which the metallic shutter side faces the SPCM. (a) Prior and posterior distribution for $f=10$. The inset shows the positive part of the distributions. (b) Posterior distribution as a function of $f$ visualised using indicated HDIs. For $f>10$, the posterior is constant. (c) Savage-Dickey ratio for null result (no light present) as function of $f$. Data points for $f\geq 30$ are fitted using a power-function.}\label{figBayesResults_v2}
\end{figure}

For our main results we used a dataset in which the black side of the shutter was facing the SPCM. In this appendix we show data for a dataset in which the (opposite) metallic side of the shutter faced the detector. Using the same methods as described in sections \ref{secMethod} and \ref{secBayes}, the measured distribution $\Delta$ is depicted in figure \ref{figCDeltaMeas_v2} after sampling for $\SI{481}{h}$ ($779220$ samples). The corresponding Bayesian fit result is presented in figure \ref{figBayesResults_v2}.\\[-0.1cm]

From figures \ref{figBayesResults_v2}(a) and \ref{figBayesResults_v2}(b), it becomes directly apparent that $p(\mu_{\Delta}^{(1)})$ has most weight for negative values of $\mu_{\Delta}$. In fact, $\text{pd}^{+}=0.026$ (probability of direction likely negative \cite{Makowskietal2019}) for this dataset and the $0.95$-HDIs for the full distribution exclude $\mu_{\Delta}^{(1)}=0$. Finally, as can be seen in figure \ref{figBayesResults_v2}(c), $r_{\text{SD}}(0)$ is below $1$ for lower values of $f$ in case it is evaluated for all MCMC samples, thus giving evidence that $\mu_{\Delta}^{(1)}\neq 0$. Hence, the data point towards an interpretation of negative light counts, which is -- of course -- non-physical.\\

Our interpretation of these results is that they stem from the so-called breakdown flash of the SPCM \cite{Lacaitaetal1993,Kurtsieferetal2001}. In single-photon avalanche photodiodes, there is a non-zero probability that photons are emitted from the diode upon detection of a photon or a dark count (a flash). There is a small probability that this flash photon reflects on the shutter and is then re-absorbed by the active area of the detector. Although such an event would happen within the dead time of the detector (approximately $\SI{50}{ns}$ for the SPCM used, whereas the flash photon is expected to return in $\SI{0.2}{ns}$ for a detector-shutter distance of $\SI{35}{mm}$), it is known that a photon absorption event towards the end of the detector dead time can lead to a so-called twilight event directly after the dead time’s completion \cite{Wareetal2007}, and we hypothesize that also earlier absorption events can lead to a similar effect. As such, this mechanism leads to an increase in counts when the shutter is closed with respect to when it is open. In case the black side of the shutter is facing the SPCM, more flash photons are absorbed instead of reflected by the shutter, which will be considered further in appendix \ref{appFlashReflection}.\\ 

\section{Model for flash photon reflection contribution}\label{appFlashReflection}
\begin{figure}[tbp]
	\centering
	\includegraphics[width=0.5\textwidth]{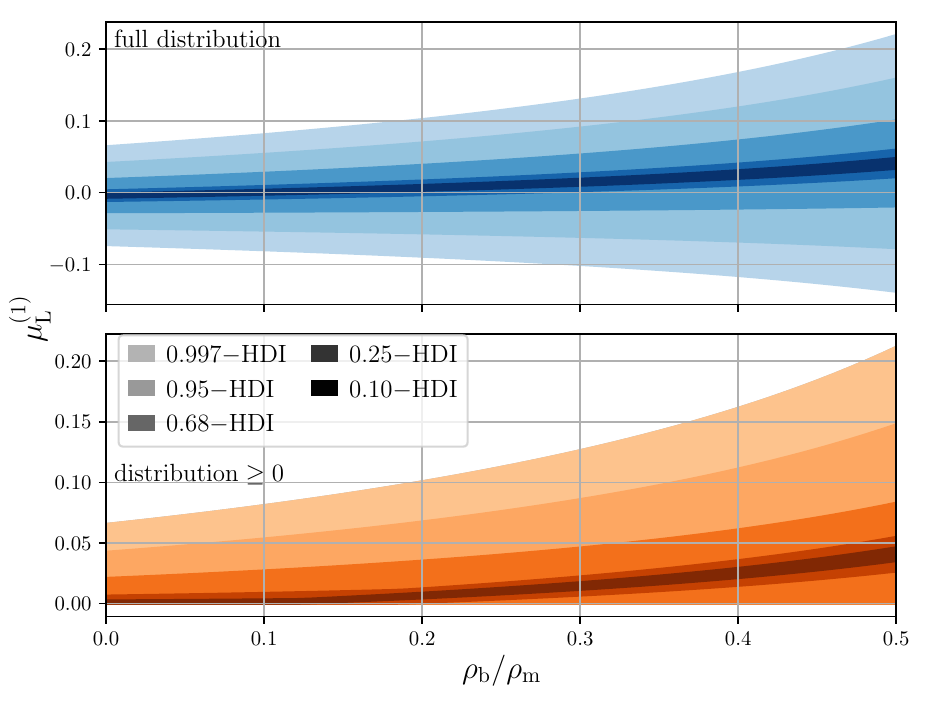}
	\caption{Posterior distribution of $\mu_{\text{L}}$ according to equation (\ref{eqCL}) as function of the relative reflection coefficient $\rho_{\text{b}}/\rho_{\text{m}}$. The top panel shows the full distribution, and the bottom panel the positive part of the distribution.}\label{figMuL}
\end{figure}

Although we find no significant deviation from $\mu_{\Delta}=0$ in our measurements with the black side of the shutter facing the SPCM, the reflected flash photons may hide a light count offset, as mentioned in section \ref{secDiscussion}. However, since we have data for two different reflective surfaces of the shutter, $k$ (black, b, and metallic, m), we can limit such an offset. Taking into account an offset, we may write
\begin{equation}
	\Delta_k\sim C_{\text{L}}-\rho_k C_{\text{F},k}.
\end{equation}
Here $C_{\text{L}}$ is the number of light counts and $\rho_k$ is the effective reflection coefficient from the shutter iris for side $k$. $C_{\text{F},k}$ is the number of flash counts emitted, where the subscript $k$ is due to the dark count fluctuations causing different count rates during the measurements for the different shutter surfaces (see figures \ref{figCDeltaMeas}(a) and \ref{figCDeltaMeas_v2}(a)). Assuming $C_{\text{F},k}\propto C_{\text{C}}$, we set $C_{\text{F,b}}\sim q C_{\text{F,m}}$, where $q=\bar{C}_{\text{C,b}}/\bar{C}_{\text{C,m}}$. Then we find
\begin{equation}\label{eqCL}
	C_{\text{L}}\sim\dfrac{\Delta_{\text{b}}-(\rho_{\text{b}}/\rho_{\text{m}})q\Delta_{\text{m}}}{1-(\rho_{\text{b}}/\rho_{\text{m}})q}.
\end{equation}
From our data we estimate $q\approx 1.05$. Substituting $\mu_{\Delta,\text{b}}^{(1)}=(-4.19\pm \SI{21.0)e-3}{cnt/s}$ and $\mu_{\Delta,\text{m}}^{(1)}=(-4.48\pm\SI{2.31)e-2}{cnt/s}$, we depict the posterior distribution for the mean of $C_{\text{L}}$, $p(\mu_{\text{L}}^{(1)})$, in figure \ref{figMuL} as function of $\rho_{\text{b}}/\rho_{\text{m}}$.\\ 
To obtain a value for $\rho_{\text{b}}/\rho_{\text{m}}$ we estimate the specular reflectivity of both shutter sides. Because the spectrum of flash photons emitted by Silicon avalanche photodiodes is in the range $750$-$\SI{1000}{nm}$ \cite{Kurtsieferetal2001}, an $\SI{852}{nm}$-laser is utilised, which is directed towards the shutter under a slight angle. The incident laser power was measured to be $\SI{14.0}{mW}$ using a power meter\footnote{Thorlabs PM$100$D with S$130$C sensor}. Positioning the power meter $\SI{20}{cm}$ away from the shutter under the angle of specular reflection, we obtain the (mostly specular) reflected laser power. From this setup, we find a reflected power of $P_{\text{r,m}}=\SI{10.0}{mW}$ for the metallic shutter side, and $P_{\text{r,b}}=\SI{3.4}{\mu W}$ for the black side. This yields $\rho_{\text{b}}/\rho_{\text{m}}\approx P_{\text{r,m}}/P_{\text{r,b}}=\SI{3.4e-4}{}$, implying the contribution of reflected flash photons in equation \ref{eqCL} is negligibly small. Therefore we can set $C_{\text{L}}=\Delta_{\text{b}}$.  

\end{document}